\documentclass[10pt,conference]{IEEEtran}
\usepackage{cite}
\usepackage{amsmath,amssymb,amsfonts}
\usepackage{algorithmic}
\usepackage{graphicx}
\usepackage{textcomp}
\usepackage{xcolor}
\usepackage{tabularx}
\usepackage{multirow}
\def\BibTeX{{\rm B\kern-.05em{\sc i\kern-.025em b}\kern-.08em
    T\kern-.1667em\lower.7ex\hbox{E}\kern-.125emX}}

\begin{document}

\title{An Empirical Study on Practicality of Specification Mining Algorithms on a Real-world Application
\thanks{University of Calgary}
}

\author{\IEEEauthorblockN{Mohammad Jafar Mashhadi}
\IEEEauthorblockA{\textit{Department of Electrical \& Computer Engineering} \\
\textit{University of Calgary}\\
Calgary, Canada \\
mohammadjafar.mashha@ucalgary.ca}
\and
\IEEEauthorblockN{Hadi Hemmati}
\IEEEauthorblockA{\textit{Department of Electrical \& Computer Engineering} \\
\textit{University of Calgary}\\
Calgary, Canada \\
hadi.hemmati@ucalgary.ca}}

\maketitle
\IEEEpeerreviewmaketitle

\begin{abstract}

Dynamic model inference techniques have been the center of many research projects recently. 
There are now multiple open source implementations of state-of-the-art algorithms, 
which provide basic abstraction and merging capabilities.
Most of these tools and algorithms have been developed with one particular application in mind, 
which is program comprehension.  
The output models can abstract away the details of the program and 
represent the software behaviour in a concise and easy to understand form. 
However, one application context that is less studied is using such inferred models for debugging, 
where the behaviour to abstract is a faulty behaviour (e.g., a set of execution traces including a failed test case).
We tried to apply some of the existing model inference techniques in a real-world industrial context to support program comprehension for debugging. 
Our initial experiments have shown many limitations both in terms of implementation as well as the algorithms. 
The paper will discuss the root cause of the failures and proposes ideas for future improvement.

\end{abstract}


\section{Introduction}

\IEEEPARstart{S}{oftware} engineers seek to understand software behaviour in all stages of development \cite{b1}. For example, during the onboarding process of new developers, 
during the verification of the program's behaviour, and during the debugging process. 

In this research, we were interested in the application of specification mining in debugging. Given that software models create a detailed but high-level and consolidated representation of the program, in the context of debugging, faulty behaviour can be abstracted in the models. This helps developers locate the root cause of bugs more effectively.
 
This research is conducted in collaboration with MicroPilot Inc.\footnote{http://micropilot.com}, a commercial Unmanned Aerial Vehicle (UAV) company located in Winnipeg, MB, Canada that develops both hardware and software. The AutoPilot software developed by this company was used as the subject of our research. It is a commercial safety-critical embedded software in C language with over 500 KLOC. 



We applied existing specification mining techniques in the area of extended finite state machine (EFSM) inference algorithms to test them in an industrial real-world setting and hopefully use them for assisting developers in the debugging process. EFSMs are like the finite state machines (FSM) with the addition of transition guards. These guards are boolean expressions that must be true prior to following a transition. \cite{efsm} We wanted to find the weaknesses of these algorithms and propose modifications and improvements as necessary. 

To achieve this goal we used the tool developed by Walkinshaw et. al. \cite{mint}
as it is a solid and promising tool that is available for free and has implemented a number of EFSM inference algorithms, including their own. We ran the regression tests for two randomly chosen bugs on the buggy version of the software and used these execution traces as the input for the tool. We tested each input with 60 configurations. 

Our attempts failed due to various reasons. We categorized the reasons into three categories: Instrumentation problems, Implementation problems, and Algorithms' shortcomings. In this paper, we dig into each category and explain the problems we faced. We discuss the attempts we made for mitigating and solving these issues in more detail. We try to find the root cause of each problem if possible and discuss different solutions for the problem in addition to the solution we used. 

\section{Background and Related work}
Program comprehension is perhaps the most well-known application of specification mining \cite{b2}, but the extent of its applications goes far beyond program comprehension. A great deal of research has been done to use mined specification models for test generation \cite{b3} \cite{b4} \cite{b5} \cite{b7}. Requirements engineering is another application of specification mining \cite{b6}. It can be used for intrusion detection via detecting anomalous behaviours of the software under test \cite{b8} as well as detecting malicious programs running on a system \cite{b9}. Model inference techniques can be used to automatically repair bugs \cite{b10}.

In general, specification mining algorithms are either dynamic or static. In static analysis approaches, techniques such as symbolic execution are used to analyze the program source code\cite{b13}. On the other hand, dynamic analysis approaches use execution traces of the programs. Execution traces can be defined as an ordered list of events that have occurred during the program execution. Event is a general term that can refer to a function call \cite{mint}, raised exception, network packet \cite{ape, synoptic}, keyboard or mouse events, and more. Events can be collected in different ways, for instance, from operating systems \cite{b8}, framework \cite{hadilogs}, or programmers' logs \cite{jiang2008automated}. 


Most previous attempts on inferring models from program execution traces focus on inferring FSMs, but there are papers that use other models such as message sequence charts \cite{damas}, LSCs \cite{lolsc}, and LTLs \cite{ltl}, as well.

We chose FSMs for two main reasons: First, The software under analysis here (AutoPilot) is heavily state-based. A majority of the reported bugs in the company's bug tracking system are related to the program performing a wrong action when it is in a certain state. The other reason is with the current state of developments of EFSM inference algorithms and tools, using state machines sounds like a more practical idea.

There are lots of studies that focus on FSM inference. For instance, the K Tails algorithm \cite{ktails} can generate FSMs from dynamic execution traces of the program. However, FSMs miss the data-related entries such as guards on the transitions. In addition, they can be potentially under-generalized and non-deterministic \cite{mint}. 
The GK Tail algorithm is an improved version of K Tails which generates EFSMs \cite{gktail} \cite{efsm}. It uses Daikon \cite{daikon} to make generalized transition guards based on the variable assignments. It is done on a per-transition basis, so it fails to incorporate the global program state in its guard inference step. 
State enhanced K Tails (SEKT) keeps track of the global program state and uses it to address this issue. \cite{sekt}
GK Tail does not prefer any merge to the others. The iterative merging is a greedy algorithm, therefore the steps that it takes may and will affect later iterations \cite{mint}.
Evidence-Driven State Merging (EDSM) \cite{edsm} tries to fix this by prioritizing merges that are supported by more evidence. It scores merge candidates and selects the best one in each step.
Walkinshaw et. al. modified EDSM by preventing state merges that lead to a non-deterministic transition situation \cite{mint}. One of their important contributions is that they developed an open source modular tool called EFSM inference tool (also known as MINT -- model inference tool) which implements a number of different inference algorithms such as K Tails, GK Tail, their own algorithm and more.

Lo et. al. empirically compared some of these inference algorithms and found out that EFSM inference techniques do not necessarily outperform FSM inference algorithms\cite{davidlo}. In another work \cite{adabu}, user input and static analysis are used to dynamically infer models for Java classes with a focus on tracking state changes. Synoptic \cite{synoptic} uses a different approach to formalize the machine inference algorithm using counterexample guided abstraction refinement\cite{synopticnphard}. 

\section{Experiment}
In this experiment, our goal is to apply some of the well-known existing algorithms for white-box dynamic model inference, on the Micropilot's AutoPilot case study, in order to assess the algorithms' applicability and find their limitations in debugging context. We assumed white-box access to the system as it is rare that someone tries to debug a program without accessing its source code. Therefore the events were in the form of function calls and the relevant state variables.

Each bug report on the company's bug reporting platform has at least one regression test associated with it. These tests are system level tests that describe flight scenarios (or partial flight scenarios) that puts the AutoPilot software into a wrong state and therefore
they generate execution traces suitable for our debugging experiments. For each bug, we reverted the AutoPilot code to the latest commit right before the bug fix. Then we used Micropilot's simulator software to run the regression tests. 

To generate the models we used the open source EFSM Inference tool. We fed our execution traces to this program and for each execution trace we ran the tool with 60 different configurations. These configurations are made of different combinations of its execution options: with and without {\verb -carefulDet } flag, with and without {\verb -daikon } flag, with 3 values of $k$ from 0 to 2, and using 5 different strategies: {\verb exhaustive }, {\verb redblue }, {\verb gktails }, {\verb noloops }, and {\verb ktails }. Further explanation of these options can be found in the tool's documentations. \footnote{https://github.com/neilwalkinshaw/mintframework/}

However, during our experiments, we faced a number of problems which we will discuss in the following sections. In the following sections we explain each problem, try to find its root cause, and we suggest solutions for fixing them and present the way we solved some of them. 

\subsection{Instrumentation phase challenges}
Previous work on dynamic model inference assume that the execution traces are readily available, but it is not the case in practice. The process of instrumenting the subject program and gathering usable execution traces has its own complications.

\subsubsection{Unmanageable number of events}
The execution traces of the AutoPilot program easily gets gigantic; hundreds of gigabytes of data can be generated in a matter of minutes. Even generating such an execution trace takes a long time let alone processing them.  That is due to the sheer complexity of the system under test. We filtered out the irrelevant data (e.g. third-party library functions) as much as possible during the instrumentation process. We used name prefixes and C++ namespaces as a heuristic to tell them apart from other functions.

\subsubsection{Sheer number of variables to be tracked}
There are a large number of variables in each function's scope. In our software under study, which was implemented in C, variables in the scope consist of both local variables (including the function input parameters) and the global variables. Keeping track of all these variables for each event is not feasible since for each function there are more than 100 Kilobytes of data accessible. We tried to keep track of the changes in the variables and only store the changes, but the result was still too large to be practically usable.

Whenever a function was called we copied a big block of memory that contained all the state variables into a buffer and computed a hash of that buffer. If it did not match the latest computed hash, it indicated a change in one or more of the variables. There are three problems with this naïve solution:

\begin{enumerate}
    \item \textbf{Nested structures and pointers:} this will be discussed more in \ref{nested_data}
    \item \textbf{Overheads:} To use this solution, a continues free block of memory having almost the same size as the working memory of the program needs to be allocated to be used as the buffer. Considering the high frequency of function calls and consequently the frequency of this data copying, this solution is costly in terms of execution time; even if the memory block was not as large.
    \item \textbf{Variable selectivity:} since we copied a whole block of memory without a context, being selective about which variables to be tracked (whitelisting) or ignoring changes in some of them (blacklisting) is devastating.  
\end{enumerate}

A solution for blacklisting can be overwriting the unwanted parts of the buffer with a fixed byte string such as 1s or 0s. But finding out the exact byte offset from which these variables start is not the most straightforward task. We tried manual whitelisting instead. We copied them into a continuous buffer (simply done using a {\verb struct }). Still, the number of CPU cycles and memory accesses required for doing it was significant and the time overhead was not decreased as much, so in the end, we abandoned this idea.

\subsubsection{Nested structures and pointers} \label{nested_data}
The aforementioned variables can have complex nested structures and multilevel indirect references that need to be flattened. For example, an array in C is stored as a pointer to the first element of the array. If we only track this pointer value, changes in the array content will be missed because a change in the content of an array does not change its memory address. This example is the simplest case. In our case, we had multiple {\verb struct }s each containing pointers to other {\verb struct }s and to one or multidimensional arrays. We ignored this problem because as mentioned before, even before flattening there are too much data to be manageable, it will unnecessarily add more complexity to an already complex problem. The size and complexity of the program determines how to address this issue. We will discuss the way we managed it in section \ref{data_selectivity}.

\subsection{Abstraction phase challenges}
In this section, we present the errors and problems we faced that were rooted in the implementation of the tool. We show that the EFSM inference tool may not be capable of processing the data generated in our scenario. We tried to preprocess our data in order to mitigate these problems (rather than modifying the tool itself as it is not the goal of this research at the moment). In some cases, we successfully managed to solve the problem, yet there are still implementation flaws that prevent our work from progressing.

\subsubsection{Data noise due to the boilerplate code}
The choice of data variables to be included in the traces affects the output greatly \cite{mint}. Some variables are not related to the high-level behaviour of the program, rather they are related to the underlying platform and can be considered boilerplate code. For example, the variables that are related to the logging (log level, log format, output file handle, etc.) do not contribute to the program behaviour and will generate unwanted noise in the data. So they should be filtered out of the execution traces.

\subsubsection{Data noise in data variables} \label{data_selectivity}
Even if we filter out the non data variables, there are still noises in the traces since not all of the data/state variables affect the functions' behaviour. Automatically selecting the ones that contribute to each functions' behaviour is a tricky problem. Daikon which is used to analyze the data variables to infer guards cannot get rid of unimportant variables when there is not enough data available to distinguish them \cite{daikon}. We faced this problem since there is usually only one regression test for each bug so we can only get one execution trace which might not be enough. MINT suggests manually handpicking the data variables \cite{mint}. So to fix these issues (Noises and also the high number of variables in the functions' scope) we asked domain expert developers to provide us with the list of functions and the variables of their interest for each bug. Then the size of the generated execution traces went under a hundred megabytes and they contained between 60K to 200K events each.

\subsubsection{Removing input redundancy} \label{data_redundancy}
The EFSM inference tool could not process the input even after these reductions in the input size. A large number of events can lead to a large prefix tree acceptor (PTA) and storing a big tree or running algorithms on it can cause different problems, e.g. Out of memory and stack overflow. We will discuss these errors more thoroughly in the next sections.

To reduce the input size furthermore, we removed similar adjacent events from the trace, when the same event happened over and over again with no change in the state variables. In these cases, we only kept one and discarded the others. In these repetitive events, some state variables might have changed but not the ones that we were tracking. This task was done using GNU's {\verb uniq } command. This process might mistakenly remove some internal transitions (transition from a state to itself) but this problem can be easily fixed by keeping two similar lines rather than just one, in this way the input size gets reduced without losing any information. Although, it might introduce internal transitions that should not actually be there. The size of the inputs was reduced up to $98\%$ in some cases. This enabled us to infer models for $25\%$ of the cases which were impossible due to errors before.

\begin{table}[!t]
    \centering
    \caption{Results of running various algorithms and configurations implemented by EFSM Inference tool on our execution traces}
    \label{tab:results_table}
    \begin{tabularx}{\columnwidth}{|l|XX|}
    \hline
    Result                   & \multicolumn{2}{c|}{Outcomes \%} \\ \hline
    Success                  &                    & $25\%$        \\ \hline
    \multirow{3}{*}{Failure} & Out of memory      & $6.67\%$      \\
                             & Stack overflow     & $28.33\%$     \\
                             & Timeout            & $40\%$        \\ \hline
    \end{tabularx}
\vspace{-0.5cm}
\end{table}

Out of all executions, 75 percent of our attempts were definite failures. (Table \ref{tab:results_table}) The tool failed to infer a model due to runtime errors. After investigating these failures we have categorized them into three classes: Timeouts, Stack overflows, and Out of Memory errors.

\subsubsection{Timeout}
The initial setup of our experiment was not using any timeouts. As a result, in some cases, it took up to 4 days to run the algorithm for only one configuration on a single input. This was far beyond acceptable, so we chose 20 minutes and later 5 minutes as the timeout value. The choice between 20 and 5 minutes did not make any difference because the tool either could finish inferring the model in less than 5 minutes or it took hours. We did not run into any input-configuration pairs that took less than an hour but more than 5 minutes to complete. But even if there were any, it is considered a failure anyway because such a long running time is more of an overhead than helping with debugging.

Among cases that could not create a model before the timeout, $79.4\%$ had the {\verb carefulDet } option on, and $76\%$ of the successful runs (created a model before timeout) had it turned off. According to the tool's documentation, enabling this option will force it to generate deterministic EFSMs to prevent over generalization. So our conclusion is that either there is a bug in that implementation or the algorithm is very time consuming. 

\subsubsection{Stack overflow}
All stack overflows happened in the functions that compute merge scores. The scores are used to find better state merges among the candidates.
Merge score is computed recursively \cite{mint} \cite{edsm} by performing a depth-first search on the FSM. Traversing a large graph recursively is likely to cause stack overflows and it did happen in our case, quite frequently. It might be worth mentioning that K Tails algorithm, which does not use merge scores, never ran into stack overflows in our tests. Recursive graph traversal algorithms can be re-implemented using loops to fix this issue.

\subsubsection{Out of memory}
What all of the failures in this class had in common was that they all were using K Tails algorithm for model inference. However, not all of the runs with this algorithm failed due to lack of memory.
This problem can have numerous root causes which we did not investigate further, but perhaps the most noticeable difference between our data and the included test data is the size of the input in terms of the number of events.

\subsection{Algorithms' shortcomings}

\subsubsection{Overgeneralized models}
Unfortunately, even the successfully created models were not quite useful for debugging purposes because the output was simple FSMs, with no data associated with the states or transitions. Even though there were no apparent implementation errors in these cases that we could point to, the algorithms did not infer transition guards. In fact, since MINT creates the data classifiers that generate transition guards before making the PTA in some cases it might intentionally skip making data classifiers for the events. For example, in our case developers chose two functions, so since function A is always followed by function B (and vice versa) the algorithm decided not to make any transition guards.

In the context of debugging, inspecting the actual state variable values that caused the bug is crucial, therefore, we consider even these successful inferences as failures.

\subsubsection{Level of details in the model}
The granularity level of the execution traces has a direct impact on the quality and usefulness of the inferred models. There is a trade-off in choosing the level of details: If low-level events are used (such as the function calls) there can be a lot of noise in the data that the inferred models misrepresent the program behaviour. On the other hand, if the events are very high level, some important details might be missed. It is worth mentioning that there have been some works on noise reduction in the low-level execution traces \cite{b2}. 

Inferring the state machine hierarchically can be an approach to address this issue. The program can be modelled as high-level components interacting with each other while each of the components can have their internal behaviour. 
\section{Limitations}

\subsection{Concurrency}
Concurrency is used in real-world applications frequently. It can be in many forms, from a multithreaded application to a network of IoT devices. In our case, multi-UAV simulation scenarios can be regarded as a high-level concurrency. Also, each UAV controller has a number of concurrent execution threads. All of these can be considered concurrent applications that each produce their own execution logs separately. To infer an accurate model, all these events and the communication between concurrent modules should be taken into consideration. 

The algorithms that we used did not take this into account so we generated models for each thread separately. But there are algorithms that focus on concurrency \cite{concurrent-fsm}. Incorporating their ideas into our research can be the goal of our future endeavours.

\subsection{Semi automated approaches}
Although full automation is useful and handy for specification mining, based on the challenges we faced we argue that occasional user inputs can help the algorithm in making better decisions in ambiguous situations. For example, in the debugging context, we argue that asking the developers to provide the set of relevant functions and state variables to a particular bug prior to generating the model is not impractical. Although user input can be useful for the algorithms, the caveat is they can become quite an overhead. Thus we plan to implement this idea and run a user study to validate the hypothesis in the future.

\subsection{Other Limitations}
The tool that we have used in this research project implements multiple EFSM inference algorithms and it was believed to be a promising and robust implementation, yet we have not explored other alternatives. A similar limitation is there for the software under study. The AutoPilot used in this study is a large-scale and safety-critical piece of software which is in use by more than 1000 customers of the company across 85+ countries, but one might argue that the study should be done on more subjects. 

As mentioned in section \ref{data_redundancy}, we removed duplicate events in order to reduce the input size and make it more manageable for the tool. But with doing so some information is lost. Some algorithms such as EDSM and its improved version --Walkinshaw et. al.'s-- rely on redundancies of this nature. These repetitions provide more ``evidence'' for some states to be considered equivalent and get merged. Removing them can prevent some states that are actually equivalent from merging and therefore resulting in under-generalized outputs.
\section{Conclusion and Future Work}
As the research matures in the domain of specification mining, it is time now to start applying the state of the art techniques on real-world large scale industrial code bases. One of the less studied but important use cases of specification mining is debugging.  In an on-going project with our industry partner, we aim to provide industrial scale solutions for specification mining --more specifically, model inference-- in the context of debugging. So far our effort has been focused on evaluating the existing tools and techniques in this domain. This paper is a summary of the challenges we faced including but not limited to implementation specific issues (e.g., stack overflow), algorithm-specific issues (e.g., over-generalization), and instrumentation challenges (e.g., making usable execution traces).
In the future, we plan to assess more tools and also improve them to handle the challenges we faced with the original works. 




\section*{Acknowledgment}
We would like to thank the three anonymous reviewers for their constructive comments.
We acknowledge the support of the Natural Sciences and Engineering Research Council of Canada (NSERC), [funding reference number CRDPJ/515254-2017].

\bibliographystyle{IEEEtran}
\bibliography{Empirical Study}

\end{document}